\documentclass[aps, preprint, showpacs, amsmath, amssymb, pra, 12pt]{revtex4}
\usepackage{graphicx}
\usepackage{dcolumn}
\usepackage{color}
\usepackage{bm}
\begin{document}
\title{Theory of single-photon transport in a single-mode waveguide coupled to a cavity containing a two-level atom}
\author{Jung-Tsung Shen}
\email{jushen@stanford.edu}
\author{Shanhui Fan}
\email{shanhui@stanford.edu}
\affiliation{Ginzton Laboratory, Stanford University, Stanford, CA
94305}
\date{\today}
\begin{abstract}
The single-photon transport in a single-mode waveguide, coupled to a cavity embedded with a two-leval atom is analyzed. The single-photon transmission and reflection amplitudes, as well as the cavity and the atom excitation amplitudes, are solved exactly via a real-space approach. It is shown that the dissipation of the cavity and of the atom respectively affects distinctively on the transport properties of the photons, and on the relative phase between the excitation amplitudes of the cavity mode and the atom.  
\end{abstract}
\pacs{42.50.-p, 03.65.Nk, 72.10.Fk} \maketitle

\section{Introduction}

Controllable single-photon transport is of central importance in quantum information processing. There have been many recent experimental~\cite{Thompson:1992a, Wallraff:2004, Birnbaum:2005, Aoki:2006, Srinivasan:2007a, Choi:2007, Dayan:2008} and theoretical~\cite{Carmichael:1985,Dalibard:1992, Tian:1992,Brun:2002, Rosenblit:2004,Bermel:2006} works probing the photon transport properties of a wavelength-scale cavity that incorporates a two- or multi-level system. These works encompass both the weak and the strong coupling regime. Traditionally, such a system is typically studied with quantum trajectory method~\cite{Dalibard:1992, Tian:1992, Brun:2002} which, as a Monte Carlo approach, is inherently stochastic~\cite{Plenio:1998, Gerry:2005}. Other approaches include employing a master equation~\cite{Carmichael:1985, Rosenblit:2004} or input-output formalism~\cite{Gardiner:1985,Domokos:2002} that assume a weak input coherent state, and usually involves uncontrolled approximations~\cite{Waks:2006}. However, the recent experiments allow us to determine the response to the input of a single injected photon. Thus, a theoretical framework that allows one to directly calculate the response of such a system to a single injected photon is valuable. 

In this article, we provide a full quantum-mechanical and deterministic approach to solve the response of this system to a single photon, yielding a wealth of information on the transport properties, as well as the effects of dissipations to the relative phase between the excitations of the cavity mode and the two-level system. Our formalism treats the problems in real space, which is particularly convenient for discussing photon transport from one space-time point to another one. Moreover, the treatment is exact, and makes no assumptions on temporal behaviors of the constituents of the system. Rather, our formalism allows direct computation of the temporal evolution of the system.

\begin{figure}[thb]
\scalebox{1}{\includegraphics{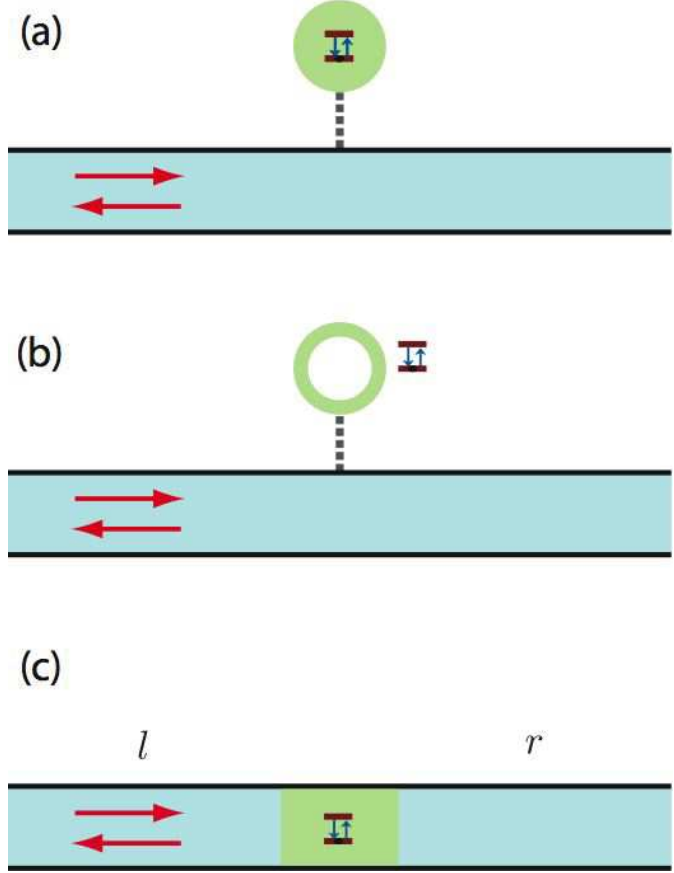}}
\caption{(color online). Schematics of the systems. A cavity interacting with a two-level atom is coupled to a single-mode waveguide in which single photons propagate in each direction. (a) Side-coupled single-mode cavity. (b) Side-coupled ring resonator. (c) Direct-coupled cases. $l$ and $r$ denote the left and right branch, respectively. The cavity is denoted by light green color, and the waveguide is denoted by the channel in light blue color. }\label{Fi:Schematics}
\end{figure}

We first describe a few configurations of the systems that is relevant to this paper. Fig.~\ref{Fi:Schematics}(a) and (b) show schematically a single-mode waveguide side-coupled to a cavity interacting with a two-level atom, where the cavity can be a single-mode microcavity (Fig.~\ref{Fi:Schematics}(a)), or a ring resonator that supports two degenerate counter-propagating modes (Fig.~\ref{Fi:Schematics}(b)).   Several notable solid-state systems are of such geometry: a superconducting quantum bit-coplanar waveguide system~\cite{Wallraff:2004}, an atom-microtoroidal resonator-waveguide geometry~\cite{Aoki:2006, Choi:2007, Srinivasan:2007a, Dayan:2008}, and a solid-state quantum dot-microdisk-waveguide system~\cite{Srinivasan:2007}. The second configuration has a single-mode cavity direct-coupled to the waveguide, as shown in Fig.~\ref{Fi:Schematics}(c). Many atomic cavity quantum electrodynamics (cavity QED) experiments use this configuration~\cite{Thompson:1992a, Birnbaum:2005}. In this paper, we will focus on systems as described in Fig.~\ref{Fi:Schematics}(a) and (c). The formalism can be generalized to treat the system in Fig.~\ref{Fi:Schematics}(b) which will be published in a separate article.

The article is organized as follows: In Sec.~\ref{SEC:Hamiltonian} the effective real-space Hamiltonian of the systems is introduced, and the exact one-photon solution is given. Sec.~\ref{SEC:NoDissipation} discusses the non-dissipative case, give the transmission spectrum, the cavity and the atomic excitation amplitude, and point out the possibility of using the system as a single-photon switch by tuning the atomic transition frequency. In Sec.~\ref{SEC:Dissipation}, we examine the effects of the dissipations in the system. Our results indicate that the dissipation of the cavity and the atom, respectively, affects the transmission and the coherence between the cavity and the atom in qualitatively different manner. Finally, in Sec.~\ref{SEC:Direct} we show that the single-photon transport of the direct-coupled case of Fig.~\ref{Fi:Schematics}(c) can be mapped into those of the side-coupled case, thus the single-photon transport properties of the two systems are related. We discuss the transmission spectrum of this case, and compare our results with experimental data of a recent circuit QED experiment.

\section{Hamiltonian and the solutions}\label{SEC:Hamiltonian}

In this section, we will derive an effective real-space Hamiltonian to describe the systems of interest. As will be shown in later sections, such a real-space representation is particularly convenient for discussing the transport properties of the photons. 

The interaction between propagating photons and a cavity with a single mode can be described by the Hamiltonian~\cite{Fan:1998b}:
\begin{equation}\label{E:Dicke}
H/\hbar = \sum_{\mathbf{k}} \omega_k c_k^{\dagger}c_k + \omega_c a^{\dagger}a + \sum_{\mathbf{k}} V_k \left(c_k^{\dagger} a + a^{\dagger}c_k\right), 
\end{equation}where $\omega_k$ is the frequency of the mode of the propagating photon field corresponding to wave vector $k$, \emph{i.e.}, the dispersion relation. $c_k^{\dagger}$ ($c_k$) is the bosonic creation (annihilation) operator of the propagating photon mode, $\omega_c$ is the resonance frequency of the cavity mode, $a^{\dagger}$ ($a$) is the bosonic creation (annihilation) operator of the cavity mode, and $V_k$ is the coupling between the cavity and the waveguide, which leads to the decay of cavity mode into the waveguide.

\begin{figure}[thb]
\scalebox{1}{\includegraphics{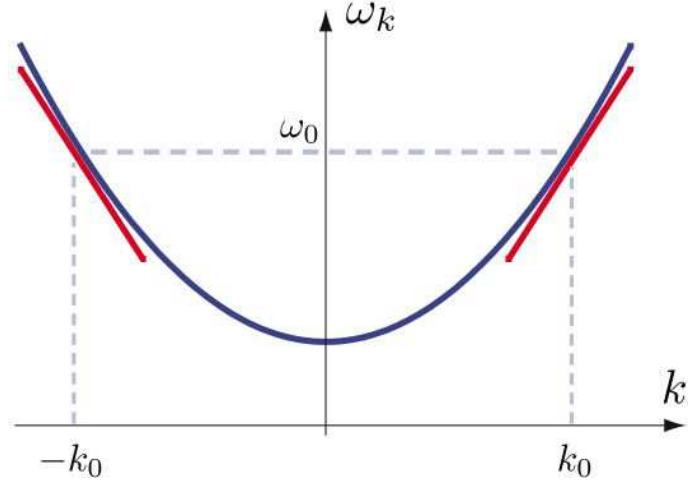}}
\caption{(color online). Linearization of the photonic dispersion of the waveguided mode. The dispersion relation, $\omega_k$, is denoted by the blue line. $\pm k_0$ are the weve vectors corresponding to an arbitrarily $\omega_0$. At the right branch around $k=k_0$, the dispersion is approximated by $\omega_{k\simeq k_0} \simeq \omega_0 + v_g (k-k_0)$, while at the left branch around $k=-k_0$, the dispersion is approximated by $\omega_{k\simeq -k_0} \simeq \omega_0 - v_g (k+k_0)$. The linearized dispersions are represented by the two red straight lines.}\label{Fi:Dispersion}
\end{figure}

We now specialize Eq.~(\ref{E:Dicke}) to the case of a single-mode waveguide. By linearizing the dispersion, the effective real-space Hamiltonian can be derived. The typical dispersion $\omega_k$ of a single-mode waveguide is shown in Fig.~\ref{Fi:Dispersion}. For an arbitrary frequency $\omega_0$ that is away from the cut-off of the dispersion, with the corresponding wave vector $\pm k_0$, one approximates $\omega_k$ around $k_0$ and $-k_0$ as 
\begin{subequations}\label{E:LinearBoth}
\begin{align}
\omega_{k\simeq k_0} &\simeq \omega_0 + v_g (k-k_0)\notag\\
&\equiv \omega_0 + v_g k_R\equiv \omega_{k_R}\quad\text{(right branch)}, \label{E:LinearRight}\\
\omega_{k\simeq -k_0} &\simeq \omega_0 - v_g (k+k_0)\notag\\
&\equiv \omega_0 - v_g k_L\equiv \omega_{k_L}\quad\text{(left branch)}, \label{E:LinearLeft}
\end{align}
\end{subequations}where the subscripts $R$ and $L$ are used to label the branch. Since we will be interested in a narrow bandwidth in vicinity of $\omega_0$, we can further extend the range of $k_R$ and $k_L$ to $(-\infty, +\infty)$. Thus, after linearizing the dispersion, one has
\begin{equation}\label{E:AfterLinearizing}
\sum_{k} \omega_k c_k^{\dagger}c_k \simeq \sum_{k_R} \omega_{k_R} c_{k_R}^{\dagger}c_{k_R}+\sum_{k_L} \omega_{k_L} c_{k_L}^{\dagger}c_{k_L}.
\end{equation} Each term on the right hand side of Eq.~(\ref{E:AfterLinearizing}) now can be easily represented in real space. By defining the Fourier transform
\begin{subequations}
\begin{align}
c_{k_R}&\equiv \int_{-\infty}^{\infty} dx\,c_R(x) e^{-i k_R x},\\
c_{k_R}^{\dagger}&= \int_{-\infty}^{\infty} dx\,c_R^{\dagger}(x) e^{+i k_R x},
\end{align}
\end{subequations}where $c_R^{\dagger}(x)$ ($c_R(x)$) creates (annihilates) a right-moving photon at $x$, the first term becomes
\begin{align}
&\sum_{k_R} \omega_{k_R} c_{k_R}^{\dagger}c_{k_R}\notag\\
=& \sum_{k_R} \left(\omega_0 + v_g k_R\right) c_{k_R}^{\dagger}c_{k_R}\notag\\
=& \sum_{k_R} \left(\omega_0 + v_g k_R\right) \iint dx dx' c_R^{\dagger}(x) c_R(x') e^{i k_R (x-x')}\notag\\
=&  \iint dx dx' c_R^{\dagger}(x) c_R(x')\left(\omega_0 - i v_g \frac{\partial}{\partial x}\right)\int_{-\infty}^{\infty} \frac{dk_R}{2\pi}e^{i k_R (x-x')}\notag\\
=&  \iint dx dx' c_R^{\dagger}(x) c_R(x')\left(\omega_0 + i v_g \frac{\partial}{\partial x'}\right)\delta(x-x')\notag\\
=& \int dx\, c_R^{\dagger}(x) \left(\omega_0 - i v_g \frac{\partial}{\partial x}\right) c_R(x).
\end{align}Similarly, by defining
\begin{subequations}
\begin{align}
c_{k_L}&\equiv \int_{-\infty}^{\infty} dx\,c_L(x) e^{-i k_L x},\\
c_{k_L}^{\dagger}&= \int_{-\infty}^{\infty} dx\,c_L^{\dagger}(x) e^{+i k_L x},
\end{align}
\end{subequations}where $c_L^{\dagger}(x)$ ($c_L(x)$) creates (annihilates) a left-moving photon at $x$, the second term becomes
\begin{align}
&\sum_{k_L} \omega_{k_L} c_{k_L}^{\dagger}c_{k_L}\notag\\
=& \int dx\, c_L^{\dagger}(x) \left(\omega_0 + i v_g \frac{\partial}{\partial x}\right) c_L(x).
\end{align}

The interaction term transforms as
\begin{align}\label{E:LinearizingInteraction}
&\sum_{k} V_k \left(c_k^{\dagger} a + a^{\dagger}c_k\right)\notag\\
=& \sum_{k_R} V_{k_R} \left(c_{k_R}^{\dagger} a + a^{\dagger}c_{k_R}\right)+\sum_{k_L} V_{k_L} \left(c_{k_L}^{\dagger} a + a^{\dagger}c_{k_L}\right)\notag\\
=& V \int \frac{dk_R}{2\pi}\left(\int dx\, c_R^{\dagger}(x) e^{+i k_R x} a+\int dx\, a^{\dagger} c_R(x) e^{- i k_R x}  \right)\notag\\
&+V\int \frac{dk_L}{2\pi}\left(\int dx\, c_R^{\dagger}(x) e^{+i k_L x} a+\int dx\, a^{\dagger} c_L(x) e^{- i k_L x}  \right)\notag\\
=& \int dx\, V\delta(x)\left(c_R^{\dagger}(x) a + a^{\dagger}c_R(x) + c_L^{\dagger}(x) a + a^{\dagger}c_L(x)\right).
\end{align}In deriving Eq.~(\ref{E:LinearizingInteraction}), $V_k$, the coupling strength between the cavity and waveguide, is assumed to be independent of $k$, and is denoted by $V$. This assumption is equivalent to the Markov approximation~\cite{Gardiner:1985}. The linearization procedure is analogous to a commonly used procedure in electronic cases, where $k_0$ is chosen as the Fermi wave vector $k_F$~\cite{ElectronLinearization}.

Having described the cavity-waveguide coupling, we now include both the atom part, as well as the interactions of the atom and the cavity with the reservoirs. Such interactions with reservoirs give rise to intrinsic dissipation~\cite{Scully:1997, Carmichael:2003}. 


The Hamiltonian of the \emph{composite} system $S\bigoplus R$ is $H\equiv H_S + H_R + H_{SR}$:
\begin{subequations}
\begin{align}
H_{S}/\hbar & \equiv \int dx \,c_R^{\dagger}(x) \left(\omega_0-i v_g \frac{\partial}{\partial x}\right) c_R(x) +  \int dx \,c_L^{\dagger}(x) \left(\omega_0+i v_g \frac{\partial}{\partial x}\right) c_L(x)\notag\\
&+\int dx \,V \delta(x) \left[a^{\dagger}c_R(x) + c_R^{\dagger}(x) a + a^{\dagger}c_L(x) + c_L^{\dagger}(x) a\right]\notag\\
&+ \omega_c a^{\dagger} a + g\left(\sigma_+ a + a^{\dagger} \sigma_{-}\right)\notag\\
&+\Omega_e a_e^{\dagger}a_e +\Omega_g a_g^{\dagger}a_g,\\ 
H_R/\hbar &\equiv H_{R_1}/\hbar + H_{R_2}/\hbar \notag\\
&\equiv \sum_j \omega_{1j} r_{1j}^{\dagger}r_{1j}+\sum_j \omega_{2j} r_{2j}^{\dagger}r_{2j},\\
H_{SR}/\hbar &\equiv H_{SR_1}/\hbar + H_{SR_2}/\hbar\notag\\
&\equiv \sum_j\left(\kappa^{*}_j r_{1j}^{\dagger}a + \kappa_j a^{\dagger} r_{1j}\right)+  \sum_j\left(\eta^{*}_j r_{2j}^{\dagger}\sigma_{-} + \eta_j \sigma_+ r_{2j}\right).
\end{align}
\end{subequations}

$H_S$ is the Hamiltonian of the system $S$ of coupled waveguide-cavity-atom. This Hamiltonian includes the atomic part, and the interaction between the atom and the cavity. $a^{\dagger}_{g}$($a^{\dagger}_{e}$) is the creation operator of the
ground (excited) state of the atom, $\sigma_{+}=a^{\dagger}_{e}
a_{g}$($\sigma_{-}=a^{\dagger}_{g} a_{e}$) is the atomic raising (lowering) ladder
operator satisfying $\sigma_{+}|n,n_c=0,-\rangle =|n,n_c=0, +\rangle$ and $\sigma_{+}|n,n_c,+\rangle =0$, where $|n, n_c, \pm\rangle \equiv |n\rangle\otimes|n_c\rangle\otimes|\pm\rangle$ describes the state of the system $S$ with $n$ propagating photons, $n_c$ photons in cavity mode, and the atom in the excited ($+$) or ground ($-$) state. $\Omega_{e}-\Omega_{g}(\equiv\Omega)$ is the atomic transition frequency. $g$ is the coupling strength between the cavity and the atom.

$H_R$ describes the reservoir, which is composed of two subsystems: $R=R_1 \bigoplus R_2$, where the cavity couples only to $R_1$, and the atom couples only to $R_2$. $R_1$ and $R_2$ are assumed to be independent. Each $R_{1}$ and $R_2$  is modeled as a collection of harmonic oscillators with frequencies $\omega_{1j}$ and $\omega_{2j}$, and with the corresponding creation (annihilation) operators $r_{1j}^{\dagger}$ ($r_{1j}$), and $r_{2j}^{\dagger}$ ($r_{2j}$), respectively.

$H_{SR}$ describes the interactions between the cavity and the atom with the reservoirs, respectively. The cavity $a^{\dagger}$ couples to the $j$th reservoir oscillator $r_{1j}$ in $R_1$ with a coupling constant $\kappa_j$, while the atom $\sigma_{+}$ couples to the $j$th reservoir oscillator $r_{2j}$ in $R_2$ with a coupling constant $\eta_j$. 

In Appendix~\ref{A:ReservoirCoupling}, we show that by incorporating the excitation amplitudes of the reservoir $R$, the effective Hamiltonian $H_{\text{eff}}$ of $S$ can be obtained and is given by:
\begin{align}\label{E:Hamiltonian}
H_{\text{eff}}/\hbar = & \int dx \left[c_R^{\dagger}(x)\left(\omega_0-i v_g\frac{\partial}{\partial x}\right)c_R(x) + c_L^{\dagger}(x)\left(\omega_0+i v_g\frac{\partial}{\partial x}\right)c_L(x)\right]\notag\\
&+\left(\omega_c -i\frac{1}{\tau_c}\right) a^\dagger a + \left(\Omega_e-i\frac{1}{\tau_a}\right) a_e^{\dagger} a_e + \Omega_g a_g^{\dagger} a_g\notag\\
&+  \int dx\, V\delta(x)\left(c_R^{\dagger}(x) a + a^{\dagger}c_R(x) + c_L^{\dagger}(x) a + a^{\dagger}c_L(x)\right)\notag\\
& + g(a\sigma_+ + a^{\dagger}\sigma_{-}),
\end{align}where $1/\tau_c \equiv \gamma_c$ and  $1/\tau_a\equiv \gamma_a$ are the dissipation rates of the cavity and the atom, respectively, due to coupling to the reservoir. We will call $H_{\text{eff}}$ as $H$ in the following for brevity. Note that $V^2/v_g$, $g$, $1/\tau_c$, and $1/\tau_a$ all have the same unit as frequency.

The temporal evolution of an arbitrary state $|\Phi(t)\rangle$ describing the system $S$ is described by the Schr\"odinger equation
\begin{equation}\label{E:Schrodinger}
i \hbar \frac{\partial}{\partial t}|\Phi(t)\rangle=H|\Phi(t)\rangle,
\end{equation} where $H$ is the effective Hamiltonian of the system in Eq.~\eqref{E:Hamiltonian}, and $|\Phi(t)\rangle$ can be expanded as
\begin{align}\label{E:State}
|\Phi(t)\rangle = & \int dx \left[\tilde{\phi}_R (x, t) c_R^{\dagger}(x) + \tilde{\phi}_L (x, t) c_L^{\dagger}(x)\right]|\emptyset\rangle\notag\\
&+ \tilde{e}_c(t) a^{\dagger}|\emptyset\rangle + \tilde{e}_a(t) \sigma_{+}|\emptyset\rangle,
\end{align}where $|\emptyset\rangle$ is the vacuum, with zero photon in both the waveguide and the cavity, and with the atom in the ground state. $\tilde{\phi}_{R/L}(x,t)$ is the single-photon wavefunction in the $R/L$ mode. $\tilde{e}_c(t)$ is the time-dependent excitation amplitude of the cavity, and $\tilde{e}_a(t)$ is the time-dependent excitation amplitude of the atom. The expansion of the state $|\Phi(t)\rangle$ in Eq.~\eqref{E:State} assumes that the atom was initially in the ground state and the cavity was empty when the incoming photon was at $-\infty$. The Schr\"odinger equation (Eq.~(\ref{E:Schrodinger})) thus gives the following set of equations of motion:
\begin{subequations}\label{E:TEoM}
\begin{align}
-i v_g\frac{\partial}{\partial x}\tilde{\phi}_R(x, t) &+ \delta(x) V \tilde{e}_c(t)+ (\omega_0+\Omega_g)\tilde{\phi}_R(x, t) = i\frac{\partial}{\partial t} \tilde{\phi}_R(x, t),\label{E:TEoM1}\\
+i v_g\frac{\partial}{\partial x}\tilde{\phi}_L(x, t) &+ \delta(x) V \tilde{e}_c(t)+ (\omega_0+\Omega_g)\tilde{\phi}_L(x, t) = i\frac{\partial}{\partial t} \tilde{\phi}_L(x, t),\label{E:TEoM2}\\
(\omega_c -i\frac{1}{\tau_c}) \tilde{e}_c(t) &+ V\left(\tilde{\phi}_R(0, t) + \tilde{\phi}_L(0, t)\right) + g \tilde{e}_a(t) + \Omega_g \tilde{e}_c(t) = i\frac{\partial}{\partial t} \tilde{e}_c(t),\label{E:TEoM3}\\
(\Omega -i\frac{1}{\tau_a}) \tilde{e}_a(t) &+ g \tilde{e}_c(t) + \Omega_g \tilde{e}_a(t)=  i\frac{\partial}{\partial t}  \tilde{e}_a(t).\label{E:TEoM4}
\end{align}
\end{subequations}For any given initial state $|\Phi(t=0)\rangle$, the dynamics of the system can be obtained straightforwardly by numerically integrating the set of equations, Eqs.~(\ref{E:TEoM}). In particular,  one could study the time-dependent transport of an arbitrary single-photon wave packet.

In the following, we concentrate on the single-photon transport of constant frequency. When $|\Phi(t)\rangle$ is an eigenstate of frequency $\epsilon$, \emph{i.e.}, $|\Phi(t)\rangle = e^{- i \epsilon t}|\epsilon^{+}\rangle$, Eq.~(\ref{E:Schrodinger}) yields the time-independent eigen equation
\begin{equation}\label{E:TimeIndependent}
H|\epsilon^+\rangle =\hbar\epsilon|\epsilon^+\rangle.
\end{equation} and the interacting steady-state solution $|\epsilon^+\rangle$ can be solved for. Here $\hbar\epsilon$ is the total energy of the coupled system $S$, with $\epsilon=\omega+\Omega_g$, and $\omega=\omega_0 + v_g k_R$.

For an input state of one-photon Fock state, the most general interacting eigenstate for the Hamiltonian of $H$ takes the following form:
\begin{align}\label{E:InteractingEigenstate}
|\epsilon^{+}\rangle = & \int dx \left[\phi_R (x) c_R^{\dagger}(x) + \phi_L (x) c_L^{\dagger}(x)\right]|\emptyset\rangle + e_c a^{\dagger}|\emptyset\rangle + e_a \sigma_{+}|\emptyset\rangle,
\end{align}where we denote the time-independent amplitudes by the corresponding untilded symbols, \emph{e.g.} $\tilde{e}_c(t) = e_c \, e^{- i \epsilon t}$, etc. The connection between the interacting eigenstate and a typical scattering experiment is described by the Lippmann-Schwinger formalism~\cite{Taylor:1972,Huang:1998,Shen:2007d}. 

The time-independent Schr\"odinger equation of Eq.~(\ref{E:TimeIndependent}) for the state $|\epsilon^+\rangle$ of Eq.~(\ref{E:InteractingEigenstate}) yields the following equations of motion:
\begin{subequations}\label{E:EoM}
\begin{align}
-i v_g\frac{\partial}{\partial x}\phi_R(x) + \delta(x) V e_c &= \left(\epsilon-\omega_0-\Omega_g\right) \phi_R(x),\label{E:EoM1}\\
+iv_g\frac{\partial}{\partial x}\phi_L(x) + \delta(x) V e_c &= \left(\epsilon-\omega_0-\Omega_g\right) \phi_L(x),\label{E:EoM2}\\
\left(\omega_c -i\frac{1}{\tau_c}\right)  e_c + V\left(\phi_R(0) + \phi_L(0)\right) + g e_a &=\left(\epsilon-\Omega_g\right) e_c,\label{E:EoM3}\\
\left(\Omega -\frac{1}{\tau_a}\right)  e_a + g e_c  &= \left(\epsilon-\Omega_g\right) e_a,\label{E:EoM4}
\end{align}
\end{subequations}with $\epsilon=\omega+\Omega_g$, and $\omega=\omega_0 + v_g k_R$.

Our aim is to solve for the transmission and reflection amplitudes for an incident photon. For this purpose, we take $\phi_R(x) = e^{i {q} x}\left(\theta(-x) + t \theta(x)\right)$, and $\phi_L(x) = r e^{-i {q} x}\theta(-x)$, where $t$ is the transmission amplitude, and $r$ is the reflection amplitude~\cite{Shen:2005, Shen:2005a}. Solving Eqs.~\eqref{E:EoM1} -\eqref{E:EoM4} for $q$, $t$, $r$, $e_c$, and $e_a$ gives:
\begin{subequations}
\begin{align}\label{E:tk}
q &= \frac{\omega-\omega_0}{v_g},\\
t & = \frac{(\omega - \Omega + i \frac{1}{\tau_a})(\omega-\omega_c + i \frac{1}{\tau_c})-g^2}{(\omega - \Omega + i \frac{1}{\tau_a})(\omega-\omega_c + i \frac{1}{\tau_c}+ i \frac{V^2}{v_g})-g^2},\label{E:t}\\
r &=\frac{-(\omega-\Omega+ i\frac{1}{\tau_a}) i \frac{V^2}{v_g}}{(\omega - \Omega + i \frac{1}{\tau_a})(\omega-\omega_c + i \frac{1}{\tau_c}+ i \frac{V^2}{v_g})-g^2},\label{E:r}\\
e_c &=\frac{(\omega-\Omega+i\frac{1}{\tau_a}) V}{(\omega - \Omega + i \frac{1}{\tau_a})(\omega-\omega_c + i \frac{1}{\tau_c}+ i \frac{V^2}{v_g})-g^2}\label{E:ec},\\
e_a &=\frac{gV}{(\omega - \Omega + i \frac{1}{\tau_a})(\omega-\omega_c + i \frac{1}{\tau_c}+ i \frac{V^2}{v_g})-g^2}\label{E:ea},
\end{align}
\end{subequations}
which are valid in both strong and weak coupling regimes.

Before we proceed, here we briefly outline some of the main features of the transmission amplitude $t$ of Eq.~(\ref{E:t}). In a waveguide-side-coupled cavity-atom system as shown in Fig.~\ref{Fi:Schematics}(a), when the cavity mode and the atom are in-tune ($\omega_c =\Omega$), the transmission spectrum is always symmetric with respect to the atomic transition frequency ($\Omega$). Moreover, the transmission spectrum has two dips with spectral separation proportional to the atom-cavity coupling, and an on-resonance photon attains local maximum. This is in contrast to the Rabi splitting peaks in a direct-coupled cavity wherein an on-resonance attains local minimum.  The transmission spectrum remains symmetric even when the cavity and the atom dissipations are present, and only becomes asymmetric when the cavity and the atom are detuned. Also, a finite cavity dissipation does not destroy the phase relation between the atom and the cavity, but the atomic dissipation does. 

Here as a side note we make a comment on a recent approach using the input-output formalism to obtain the transmission and reflection amplitudes. In Ref.~[\onlinecite{Waks:2006}], in order to linearize the nonlinear term $i g \sigma_z(t) a(t)$ in the  equations of motion of the Heisenberg operators (See Eq.~(2) in Ref.~[\onlinecite{Waks:2006}]), the two-level atom is approximated to be in the ground state all the time, such that $\sigma_z (t)$ is subsequently substituted with $-1$. In fact, this approximation is non-physical, as can be seen from the exact solutions here, since the atom is strongly excited even with a single photon, in the absence of strong dissipation, which is the case of interest here. Such excitation is crucially important for using the atom to control the photon transport.  
Instead, one can show that the equations of motion in the Heisenberg picture for the operators can give rise to a set of exact equations of motion of the amplitudes that is closed, provided we sandwich the operators with $|\epsilon^+\rangle$ and vacuum state $\langle \emptyset|$. This set of equations has solutions identical to Eq.~(\ref{E:t})-(\ref{E:ea}). In particular, the nonlinear term $i g \sigma_z(t) a(t)$ gives the matrix element $\langle \emptyset| i g\sigma_z(t) a(t) |\epsilon^+\rangle = -i g e_c e^{-i \omega t}$, which, after dividing the common factor $-i e^{-i \omega t}$ on both sides of Eq.~(2) in Ref.~[\onlinecite{Waks:2006}], gives the $g e_c$ term as in Eq.~(\ref{E:EoM4}). In doing so, one can then derive all the amplitudes, $e_c$ etc., and hence determines $|\epsilon^+\rangle.$ The single-photon transport is exactly solvable in both Schr\"{o}dinger picture and Heisenberg picture.

Below we investigate the effects of dissipations on the transmission spectrum. We start by discussing the non-dissipative case, followed by examining the dissipations of the cavity and of the atom. For each case, we will focus on the changes due to dissipations on the local maxima and minima determined from $d |t|^2/d\omega \equiv 0$.

\section{Non-dissipative case ($1/\tau_c=1/\tau_a=0$):}\label{SEC:NoDissipation}

\begin{figure}[thb]
\scalebox{1}{\includegraphics[width=\columnwidth]{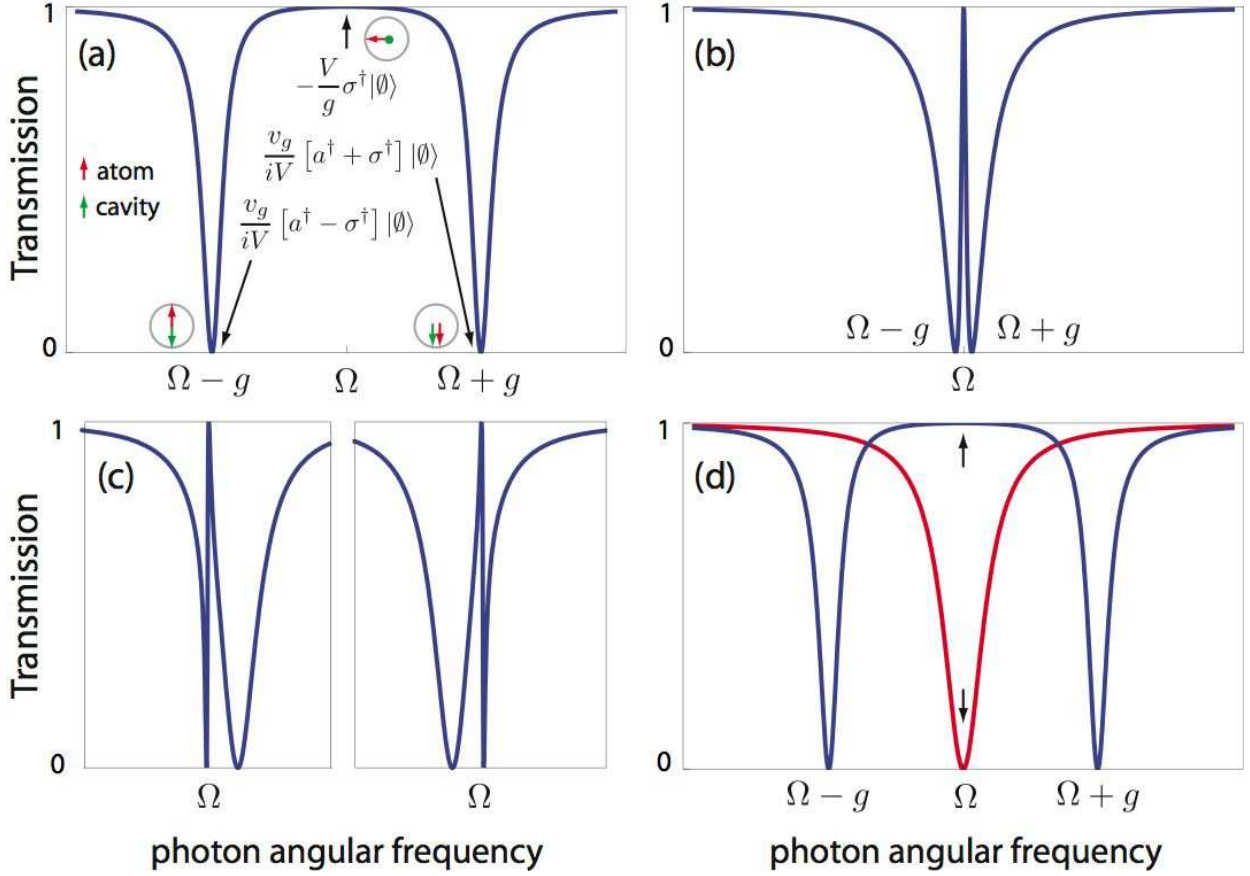}}
\caption{(color online). Single-photon transmission spectrum for non-dissipative case. (a) In-tune, non-dissipative case ($\Omega=\omega_c$; $1/\tau_c=1/\tau_a=0$).  At $\omega=\Omega$, the transmission is 1, and at $\omega=\Omega\pm g$, the transmission is 0. $g=0.5\Omega$, $V^2/v_g=0.09\Omega$ are used for plotting the spectra. The conclusions however are independent of the choice of the numerical values. (b) When $g$ is small, the transmission peak becomes narrow, analogous to EIT.  $g=0.03\Omega$, $V^2/v_g=0.09\Omega$. (c) Atom-cavity detuned, non-dissipative case ($\Omega\neq\omega_c$; $1/\tau_c=1/\tau_a=0$). Left:  $\omega_c = 1.1 \Omega$, $V^2/v_g=0.09\Omega$. Right: $\omega_c = 0.9 \Omega$, $V^2/v_g=0.09\Omega$. The transmission is always 1 at $\omega=\Omega$. (d) Single-photon switching. The red curve is the transmission spectrum when the atom is far detuned ($|\Omega - \omega_c| V^2/v_g \gg g^2$), or is not present. The transmission of an on-resonance photon with $\omega=\omega_c$ thus changes from 1 to essentially 0 by tuning the atomic transition frequency.}\label{Fi:NoDissipations}
\end{figure}

\noindent(1)\emph{Atom-cavity in tune} ($\Omega=\omega_c$): The transmission spectrum for this case is shown in Fig.~\ref{Fi:NoDissipations}(a). The spectrum exhibits a maximum at $\omega=\Omega$. At $\omega=\Omega$, the on-resonance photon completely transmits ($t=1$),  and the cavity is not excited ($e_c=0$),  but only the atom is excited ($e_a=-V/g$). We note that, in contrast, when only the atom or the cavity is present, an on-resonance photon is completely reflected~\cite{Shen:2005}. 
The transmission spectrum shows two minima at $\omega=\Omega\pm g$, which correspond to the Rabi-splitted frequencies. 
At $\omega=\Omega \pm g$, the photon is completely reflected ($r=-1$). Also, $e_c = v_g/(i V) = \pm e_a$, thus the cavity excitation and the atom excitation are of equal amplitude and either in-phase or completely out of phase with each other.  The full-width at half-minimum ($|t|^2=1/2$) of each dip is exactly $V^2/v_g$, independent of the atom-photon coupling constant $g$. (In comparison, for a side-coupled cavity with no atom embedded, the cavity decay rate is also $V^2/v_g$). The spectral distance between the two local minima is $2 g$. When $g$ is small, the transmission peak thus becomes very narrow, exhibiting a spectrum that is analogous to electromagnetically induced transparency phenomena~\cite{Harris:1997, Waks:2006}, as shown in Fig.~\ref{Fi:NoDissipations}(b). 

\noindent(2)\emph{Atom-cavity detuned} ($\Omega\neq\omega_c$):  When the photon is on-resonant with the atom($\omega=\Omega$), the transmission amplitude is always 1, regardless of the detuning between the atom and the cavity, as shown in Fig.~\ref{Fi:NoDissipations}(c). On the other hand, when the atom is far detuned from the cavity resonance frequency ($|\Omega - \omega_c| V^2/v_g \gg g^2$), the single photon transmission spectrum has dips down to zero at the cavity frequency $\omega \simeq \omega_c$. This feature could be exploited to achieve a fast single-photon switch: for an incoming photon with frequency $\omega = \omega_c$, the transmission is 1 when the atom is in-tuned with the cavity ($\Omega = \omega_c$); while the transmission is essentially 0 when the atom is far detuned. Thus, by tuning the transition frequency of the atom, the single-photon transport can be regulated and the setup acts as a single-photon switch, as shown by Fig.~\ref{Fi:NoDissipations}(d). This effect was pointed out in Ref.~[\onlinecite{Waks:2006}], here we give an exact derivation of this effect.

\section{The effects of dissipations}\label{SEC:Dissipation}

The unavoidable intrinsic dissipative processes in the system always result in the leakage of photons into non-waveguided degrees of freedom. In general, the dissipations affect the transmission properties, and change the phase relation between the excitation amplitudes of the cavity mode and the atom. However, as we show below, the dissipation of the cavity and of the atom affect these properties rather differently. 

\begin{figure}[thb]
\scalebox{1}{\includegraphics[width=\columnwidth]{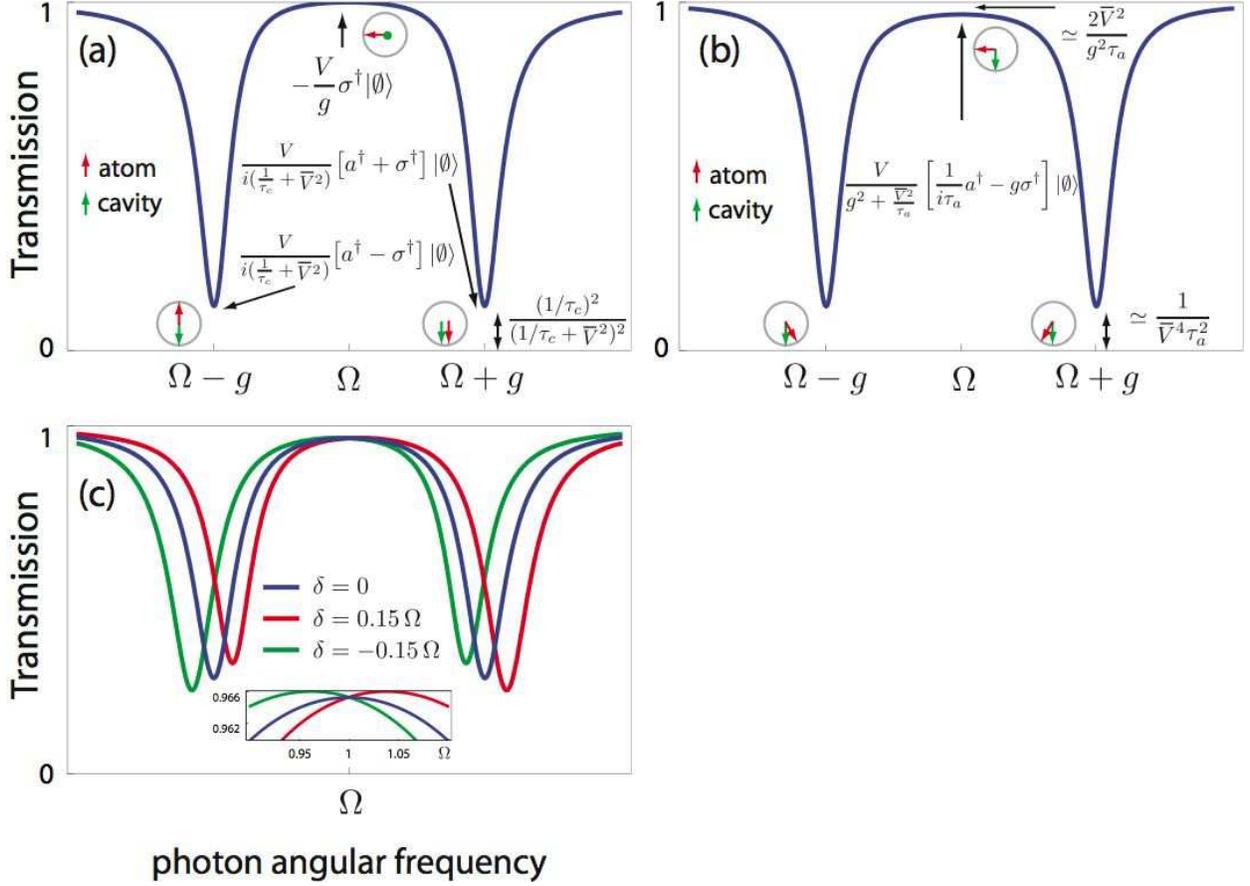}}
\caption{(color online). Single-photon transmission spectrum for dissipative cases.  (a) In-tune, dissipative cavity case ($\Omega=\omega_c$; $1/\tau_c=0.05\Omega$). The transmission is 1 at $\omega=\Omega$, and is  $\simeq v_g^2/(V^4 \tau_c^2)$ at $\omega=\Omega\pm g$. (b) Detuned, dissipative atom case ($\Omega=\omega_c$; $1/\tau_a =0.05\Omega$). The transmission is $\simeq 2(V^2/v_g)/(g^2 \tau_a)$ at $\omega=\Omega$, and is $\simeq v_g^2/(V^4 \tau_a^2)$ at $\omega=\omega\pm g$. Also shown are the (cavity $+$ atom) component of the eigenstate, and the relative phase denoted by the green and red arrows. (c) Detuned, dissipative cavity and atom case  ($\Omega \neq\omega_c$; $1/\tau_c=1/\tau_a=0.05\Omega$). The spectrum becomes asymmetric when $\Omega\neq\omega_c$, and the local maximum is no longer located at $\omega=\Omega$. For small dissipations, however, the transmission at $\omega=\Omega$ is insensitive to the detuning, $\delta$, as shown in the inset. $g=0.5\Omega$, $V^2/v_g=0.09\Omega$ are used for plotting the spectra. The conclusions however are independent of the choice of the numerical values. $\bar{V}^2\equiv V^2/v_g$.}\label{Fi:AllCases}
\end{figure}

\noindent (1)\emph{Atom-cavity in-tune, dissipative cavity} ($\Omega=\omega_c$; $1/\tau_a=0$ and $1/\tau_c\neq 0$): The transmission spectrum is shown in Fig.~\ref{Fi:AllCases}(a). The overall transmission spectrum remains symmetric with respect to $\omega=\Omega$. The locations of the local maximum $\omega=\Omega$, and the local minima $\omega=\Omega\pm g$, are independent of cavity dissipation $1/\tau_c$ and are the same as in the dissipationless case. At $\omega=\Omega$, one still has $t=1$, $r=e_c=0$, and $e_a=-V/g$. Remarkably, a photon at this frequency still transmits perfectly and the cavity is not excited, even with the presence of the cavity dissipation $1/\tau_c$. One can understand this result from the non-dissipative case: since the cavity is not excited for a photon at $\omega=\Omega$, the cavity dissipation makes no effect on the transmission. 
At $\omega=\Omega\pm g$, one has 
\begin{equation}
t =\frac{\frac{1}{\tau_c}}{\frac{1}{\tau_c}+\frac{V^2}{v_g}}
\end{equation} for both frequencies. $|t|^2$ approximates $\frac{1}{(\tau_c V^2/v_g)^2}$ when the decay rate is small such that $1/\tau_c \ll V^2/v_g$, \emph{i.e.}, when the coupling rate between the waveguide and the cavity is larger compared with the cavity dissipation rate. Moreover, at $\omega=\Omega\pm g$, $e_c  = -i V/(1/\tau_c + V^2/v_g)= \pm e_a$. Thus, while the excitation amplitudes are reduced from the previous case, both cavity and atomic excitations are still either in-phase or out-of-phase with each other with the presence of cavity dissipation and has the same amplitude. Thus, the cavity on one hand largely insulates the atom from decohering interactions with external environment~\cite{Mabuchi:2002a}, on the other hand its own dissipation does not affect either the perfect transmission of an on-resonace photon, or the phase relation between the cavity and the atom at the Rabi-splitted frequencies ($\omega=\Omega\pm g$). When the dissipation is small, the full-width at half-maxium ($|t|^2=(1+|t_{k=\Omega\pm g}|^2)/2$) of each dip is exactly $V^2/v_g + 1/\tau_c$, independent of the atom-photon coupling constant $g$.

\noindent (2)\emph{Atom-cavity in-tune, dissipative atom case} ($\Omega=\omega_c$; $1/\tau_c=0$ and $1/\tau_a\neq 0$): The transmission spectrum is plotted in Fig.~\ref{Fi:AllCases}(b). The overall transmission spectrum remains symmetric with respect to $\omega=\Omega$. The local maximum is still located at $\omega=\Omega$, independent of atom dissipation $1/\tau_a$ and is the same as in the dissipationless case. The local minima $\omega_{\pm}$, accurate up to the third order of $\frac{1}{\tau_a}$, are $\omega_{\pm}\simeq\Omega\pm \left(g- \frac{v_g}{2 g V^2}\left(\frac{1}{\tau_a}\right)^3\right)$. 

At $\omega=\Omega$, one has 
\begin{align}
t&=\frac{g^2}{g^2+\frac{V^2/v_g}{\tau_a}},\notag\\
r& = -\frac{\frac{V^2/v_g}{\tau_a}}{g^2 + \frac{V^2/v_g}{\tau_a}},\notag\\
e_c &= -i \frac{V/\tau_a}{g^2 + \frac{V^2/v_g}{\tau_a}} \neq 0,\notag\\
e_a &= -\frac{g V}{g^2+\frac{V^2/v_g}{\tau_a}}.
\end{align} Comparing with the dissipative cavity case above, we thus see that the atom dissipation has a stronger effect on the transmission of an on-resonance photon than the cavity dissipation does.  
The relative phase $\phi$ between the cavity excitation and the atom excitation is always $\pi/2$ for any value of non-zero atomic dissipation. 

Accurate up to second order of $\frac{1}{\tau_a}$, $\omega_{+}\simeq \Omega+g$, and the transmission amplitude is
\begin{equation}\label{E:tkbig}
t =\frac{i\frac{1}{\tau_a}g}{-\frac{V^2/v_g}{\tau_a} + i(\frac{1}{\tau_a} + \frac{V^2}{v_g})g},
\end{equation}
thus $|t|^2 \simeq \left(\frac{v_g}{V^2 \tau_a}\right)^2$ when the dissipation is small. $|r|^2 \simeq 1-2 v_g/(V^2 \tau_a)$ at the same limit. Moreover, the cavity excitation and the atom excitation are no longer in-phase ($e_c \neq e_a$), even at the order of $1/\tau_a$. The relative phase $\phi$ between the cavity excitation and the atom excitation is given by $\tan\phi = 1/(g\tau_a)$. Similarly, accurate up to second order of $\frac{1}{\tau_a}$, $\omega_{-}\simeq \Omega - g$, and the transmission amplitude is
\begin{equation}
t =\frac{i\frac{1}{\tau_a}g}{\frac{V^2/v_g}{\tau_a} + i(\frac{1}{\tau_a} + \frac{V^2}{v_g})g},
\end{equation}which is complex conjugate of $t$ at $\omega=\Omega+g$ in Eq.~(\ref{E:tkbig}).
The cavity excitation and the atom excitation are no longer completely out-phase ($e_c \neq -e_a$) even at the order of $1/\tau_a$. The relative phase $\phi$ between the cavity excitation and the atom excitation is $\tan\phi = -1/(g\tau_a)$. Nonetheless, at both Rabi-splitted frequencies, $|e_c|^2 \simeq |e_a|^2 = \left(\frac{v_g}{V}\right)^2\left(1-2\frac{v_g}{V^2 \tau_a}\right)$. The full-width at half-maxium ($|t|^2=(|t_{\omega=\omega_{\pm}}|^2+ |t_{\omega=\Omega}|^2)/2$) of each dip is  $V^2/v_g + 1/\tau_a$, independent of the atom-photon coupling constant $g$. 

\noindent (3)\emph{Atom-cavity in-tune, dissipative cavity and atom case} ($\Omega=\omega_c$; $1/\tau_c$, $1/\tau_a\neq 0$): The local maximum is still located at $\omega=\Omega$ exactly, while the Rabi-splitted frequencies are shifted to $\omega_{\pm}\simeq \Omega\pm \left(g+ \frac{1}{2 g \tau_c \tau_a}\right)$. The transmission spectrum remains symmetric with respect to $\omega=\Omega$. At $\omega=\Omega$, 
\begin{equation}
t = \frac{\frac{1}{\tau_c}\frac{1}{\tau_a}+g^2}{\frac{1}{\tau_c}\frac{1}{\tau_a} + \frac{V^2/v_g}{\tau_a}+g^2},
\end{equation}
thus $|t|^2$ approximates $1-2\frac{V^2/v_g}{g^2\tau_a}$, when $1/\tau_a < g, V^2/v_g$, and $1/\tau_c$, and is independent of cavity decay rate $1/\tau_c$. $|r|^2 \simeq \left(\frac{V^2/v_g}{g^2 \tau_a}\right)^2$ at this limit. The cavity is slightly excited with $|e_c|^2\simeq V^2/(g^4 \tau_a^2)$. The relative phase $\phi$ between the cavity excitation and the atom excitation is always $\pi/2$ for any value of non-zero atomic dissipation. At $\omega=\omega_{+}$, $|t|^2\simeq (v_g^2/V^4)(1/\tau_c + 1/\tau_a)^2$. The cavity excitation and the atom excitation are not in-phase at the order of $1/\tau_a$. Similarly, at $\omega=\omega_{-}$, $|t|^2\simeq (v_g^2/V^4)(1/\tau_c + 1/\tau_a)^2$, and the cavity excitation and the atom excitation are not completely out-phase even at the order of $1/\tau_a$. Also, the transmission amplitudes at two local minima are complex conjugate to each other, $t(\omega_{+}) = {t}^{*}(\omega_{-}).$

\noindent (4)\emph{Detuned, dissipative cavity and atom case} ($\Omega \neq\omega_c$; $1/\tau_c$, $1/\tau_a\neq 0$): When the cavity and the atom are in-tune, the single-photon transmission spectrum is always symmetric with respect to $\omega=\Omega$, regardless of the dissipations. When the atomic transition frequency $\Omega$ is detuned slightly away from the cavity frequency $\omega_c$, the transmission spectrum becomes asymmetric. The frequency of the local maximum is now located between $\Omega$ and $\omega_c$. Remarkably, there exists an optimal frequency for the incident single photons such that at which frequency the magnitude of the transmission is \emph{insensitive} to the frequency detuning $\delta (\equiv \Omega -\omega_c)$. This ``cavity protected'' optimal frequency can be obtained by expanding $d |t|^2/d\delta=0$ with respect to $\delta$ and numerically solving the resultant polynomial equation. When the cavity and atom dissipations are small, retain only the first order term is enough, and the optimal frequency is essential at $\omega=\Omega$, as demonstrated in Fig.~\ref{Fi:AllCases}(c). Such optimal frequency could be useful to stabilize the single-photon transmission when the cavity frequency experiences drifting. Also, in the far-detuned case, the atom is essentially decoupled from the cavity, thus the transmission properties at $\omega \simeq \omega_c$ are determined by the cavity only.

\section{Direct-coupling case}\label{SEC:Direct}

\subsection*{V1. Direct-coupling vs side-coupling}
Another configuration often employed in photon transport experiments is to directly place the cavity in the waveguide, as shown in Fig.~\ref{Fi:Schematics}(c). In this case, the transmitted photons in the forward direction can only do so by tunneling in and out of the cavity. In comparison, the reflected photons in the side-coupled case also has only contribution from the cavity. Thus, one anticipates that the \emph{transmission} amplitude of the direct-coupled case is related to the \emph{reflection} amplitude of the side-coupled case. We show in the following that it is indeed so.

\begin{figure}[thb]
\scalebox{1}{\includegraphics{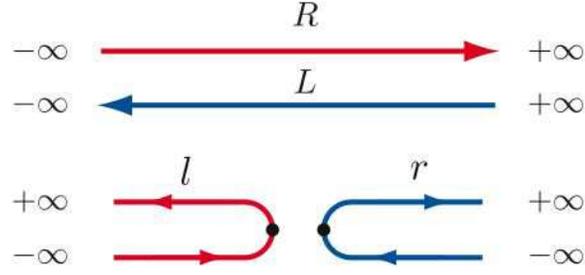}}
\caption{(color online). Folding of the waveguiding paths for photons for the side-coupled case and the direct-coupled case. The $l$-branch mode of the direct-coupled case is folded from the $R$-mode of the side-coupled case, and the $r$-branch mode is folded from the $L$-mode. The orientations of $l$- and $r$-branch are such that the incident waves come from $x=-\infty$, and the outgoing waves runs toward $x=+\infty$. The black dots indicate a phase shift in the reflected paths.}\label{Fi:Folding}
\end{figure}

To describe the direct-coupled case in Fig.~\ref{Fi:Schematics}(c), we first construct the Hamiltonian for the system, which involves the fields in the left ($l$) and right ($r$) branches, and their interaction via the cavity, as shown in Fig.~\ref{Fi:Folding}. For the left branch, we define a field $c_l^{\dagger}(x)$ such that $c_l^{\dagger}(x<0)$ describes a photon that is moving to the right at $-|x|$ in the left branch, and $c_l^{\dagger}(x>0)$ describes a photon that is moving to the left at $-|x|$ in the left branch. Thus, $c_l^{\dagger}(x<0)$ describes an incoming photon and $c_l^{\dagger}(x>0)$ describes an outgoing photon in the left branch. In order to take into account the phase shift that occurs during the reflection at the end of the waveguide, we write the following Hamiltonian
\begin{equation}\label{E:LeftBranch}
H_l/\hbar= \int_{-\infty}^{+\infty} dx\, c_l^{\dagger}(x)\left(-i v_g\frac{\partial}{\partial x}\right)c_l(x)- \int_{-\infty}^{+\infty} dx\, v_g \varphi \frac{\partial f}{\partial x} c_l^{\dagger}(x) c_l(x).
\end{equation} Here $\varphi$ is the phase shift due to reflection, as shown below. $f(x)$ is a switch-on function with the general property that $\lim_{x\rightarrow -\infty}f(x)= 0$ and $\lim_{x\rightarrow +\infty}f(x) = 1$ in a short spatial extent, otherwise the specific form is unimportant. For computational purpose, one can take, for example, $f(x)=\frac{1}{1+e^{-x/a}}$, where the spatial turning range is controlled by $a$. 

The Hamiltonian in Eq.~(\ref{E:LeftBranch}) is in fact equivalent to a free Hamiltonian via the 
the following canonical (gauge) transformation:
\begin{equation}\label{E:lToR}
c_{l}(x) \equiv e^{i\varphi f(x)} c_R(x),
\end{equation}\emph{i.e.}, the $l$-branch mode of the direct-coupled case is folded from the $R$-mode of the side-coupled case. Eq.~(\ref{E:LeftBranch}) becomes
\begin{equation}\label{E:HamiltonianltoR}
H_R/\hbar\equiv \int dx\, c_R^{\dagger}(x)\left(-i v_g\frac{\partial}{\partial x}\right)c_R(x).
\end{equation} The corresponding eigen wavefunctions of $H_l$ (Eq.~\eqref{E:LeftBranch}) and of $H_R$ (Eq.~\eqref{E:HamiltonianltoR}) transform to each other according to
\begin{equation}
\int dx\, e^{i {q} x} c_R^{\dagger}(x)|\emptyset\rangle =\int dx\, e^{i {q} x} e^{i\varphi f(x)} c_l^{\dagger}(x)|\emptyset\rangle,
\end{equation}Thus, in the left branch, for an incoming wave $e^{i {q} x}$ at $x\rightarrow -\infty$, the outgoing wave (\emph{i.e.}, reflected wave) acquires a phase $\varphi$ to become $e^{i {q} x + i\varphi}$ at $x\rightarrow +\infty$. 

Similarly, for the right ($r$) branch in Fig.~\ref{Fi:Folding}, we define a field $c_r^{\dagger}(x)$ such that $c_r^{\dagger}(x>0)$ describes a photon that is moving to the right at $+|x|$ in the right branch, and $c_r^{\dagger}(x<0)$ describes a photon that is moving to the left at $+|x|$ in the right branch. Thus, $c_r^{\dagger}(x<0)$ describes an incoming photon and $c_r^{\dagger}(x>0)$ describes an outgoing photon in the right branch. The field in the right branch is described by the Hamiltonian
\begin{equation}\label{E:RightBranch}
H_r/\hbar\equiv\int_{-\infty}^{+\infty} dx\, c_r^{\dagger}(x)\left(- v_g i\frac{\partial}{\partial x}\right)c_r(x)- \int_{-\infty}^{+\infty} dx\, v_g \varphi \frac{\partial f}{\partial x} c_r^{\dagger}(x) c_r(x).
\end{equation}Via the canonical transformation
\begin{equation}\label{E:rToL}
c_{r}(x) \equiv e^{i\varphi f(x)} c_L(-x)
\end{equation} \emph{i.e.}, the $r$-branch mode of the direct-coupled case is folded from the $L$-mode of the side-coupled case, the Hamiltonian $H_r$ of Eq.~\eqref{E:RightBranch} can be transformed into a free Hamiltonian
\begin{equation}
H_L/\hbar\equiv \int dx\, c_L^{\dagger}(x)\left(+i v_g \frac{\partial}{\partial x}\right)c_L(x),
\end{equation}with the wavefunction transformed as
\begin{equation}
\int dx\, e^{-i {q} x} c_L^{\dagger}(x)|\emptyset\rangle = \int dx\, e^{-i {q} x} e^{i\varphi f(-x)}c_r^{\dagger}(-x)|\emptyset\rangle=\int dx\, e^{+i {q} x} e^{i\varphi f(x)}c_r^{\dagger}(x)|\emptyset\rangle.
\end{equation}

When the cavity is present to allow the tunneling between the $l$ and $r$ branches, the Hamiltonian is
\begin{align}
H/\hbar=&\int dx\, c_l^{\dagger}(x)\left(-i v_g\frac{\partial}{\partial x}\right)c_l(x)- \int dx\, v_g\varphi \frac{\partial f}{\partial x} c_l^{\dagger}(x) c_l(x)\notag\\
&+\int dx\, c_r^{\dagger}(x)\left(-v_g i\frac{\partial}{\partial x}\right)c_r(x)- \int dx\, v_g \varphi \frac{\partial f}{\partial x} c_r^{\dagger}(x) c_r(x)\notag\\
&+ \int dx\, V\delta(x)\left(c_l(x) a^{\dagger}+ a c_l^{\dagger}(x) + c_r(x) a^{\dagger}+ a c_r^{\dagger}(x)\right) \notag\\
&+ \omega_c a^{\dagger} a + \Omega a_e^{\dagger} a_e+  g(a \sigma_{+}+\sigma_{-} a^{\dagger}),
\end{align}which, using Eq.~(\ref{E:lToR}) and (\ref{E:rToL}), can be written as:
\begin{align}\label{E:HamiltonianWithPhase}
H/\hbar=&\int dx\, c_R^{\dagger}(x)\left(-i v_g\frac{\partial}{\partial x}\right)c_R(x)+\int dx\, c_L^{\dagger}(x)\left(+i v_g\frac{\partial}{\partial x}\right)c_L(x)\notag\\
&+ \int dx\, V\delta(x)\left(c_R(x) a^{\dagger} e^{i\varphi f(0)}+ a c_R^{\dagger}(x) e^{-i\varphi f(0)}+ c_L(x) a^{\dagger} e^{i\varphi f(0)}+ a c_L^{\dagger}(x)e^{-i\varphi f(0)}\right) \notag\\
&+ \omega_c a^{\dagger} a + \Omega a_e^{\dagger} a_e+  g(a \sigma_{+}+\sigma_{-} a^{\dagger}),
\end{align} similar to the Hamiltonian of Eq.~(\ref{E:Hamiltonian}) but with the additional phase terms in the cavity-waveguide coupling terms.

It can be shown straightforwardly that the single-photon transmission and reflection amplitudes for systems described by the Hamiltonian of Eq.~(\ref{E:Hamiltonian}) and of Eq.~(\ref{E:HamiltonianWithPhase}) respectively are identical and are both independent of $\varphi$ and $f(0)$. Only the cavity and atomic excitation amplitudes depend upon $\varphi$ and $f(0)$. The relations between each eigenfunctions are given by
 $\phi_{l}(x)=\phi_R (x) e^{i\varphi f(x)} = e^{i {q} x}\left(\theta(-x)+ t \theta(x)\right)e^{i\varphi f(x)}$, and $\phi_{r}(x)=\phi_L (-x)e^{i\varphi f(x)}= r e^{i {q} x}\theta(x) e^{i\varphi f(x)}$, \emph{i.e.}, in the direct-coupled case the \emph{reflection} amplitude is $t e^{i\varphi}$ and the \emph{transmission} amplitude is $r e^{i\varphi}$, where $t$ and $r$ are the transmission and reflection amplitudes of the side-coupled case discussed above (Eq.~(\ref{E:t}) and (\ref{E:r})). This relation establishes the intuitive physical picture at the beginning of this section.

\subsection*{V2. Comparison with the experiment}

\begin{figure}[thb]
\scalebox{1}{\includegraphics[width=\columnwidth]{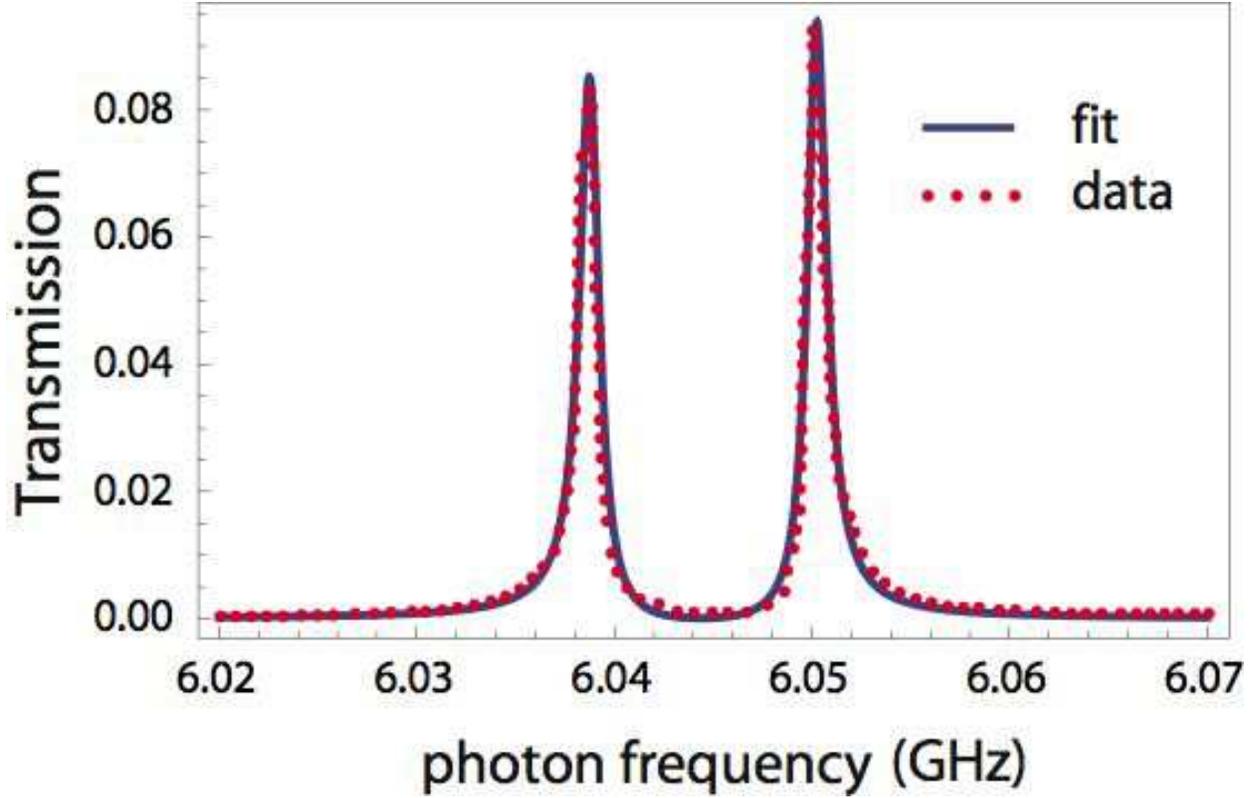}}
\caption{(color online). The fitting of the transmission spectrum in Ref.~\cite{Wallraff:2004}. The extracted experimental data are denoted by the red dots, while the blue curve indicates the fitting using Eq.~(\ref{E:r}), with the following parameters: $\omega_c/2\pi=6.0446$ GHz, $\Omega/2\pi=6.0444$ GHz, $V^2/v_g/2\pi=0.361$ MHz, $\gamma_c=0$, $\gamma_a/2\pi=0.86$ MHz, and $g/2\pi=5.73$ MHz (\emph{i.e.}, the Rabi frequency is $2g/2\pi = 11.46$ MHz). These values are very close to those of experimental fitting in Ref.~\cite{Wallraff:2004}, except where an in-tune condition ($\omega_c=\Omega$) is assumed.}\label{Fi:Fitting}
\end{figure}

To demonstrate the validity of our approach, we apply our formalism to the transmission spectrum of single photons in a solid-state circuit QED system~\cite{Wallraff:2004}, where a superconducting Josephson junction qubit is embedded in a cavity capacitively direct-coupled to a transmission line waveguide (Fig.~\ref{Fi:Schematics}(c)), thus the transmission amplitude is described by $r e^{i\varphi}$ (Eq.~(\ref{E:r})). Fig.~\ref{Fi:Fitting} shows the results, where it is clearly seen that the experimental data can be fitted extremely well by $|r|^2$~\cite{NoisyData}. Importantly, our results indicate that the asymmetry in the two transmission peaks at the Rabi frequencies is likely due to very slight frequency detuning between the cavity and the qubit.



\section{summary}\label{SEC:Summary}

Dissipation and decoherence processes limit the performances of quantum information processing devices. A thorough understanding on such processes therefore are of vital importance in the realistic implementations of any such quantum devices. We have shown the impacts of the dissipations on the entanglement in a waveguide-cavity-atom system. Here we make some final remarks on our approach.  Our approach could be generalized to take into account of other atomic degrees of freedom, such as multi-level atomic system, as well as the motion and the polarizability of the atom. This allows the alternative treatment of a number of interesting one-dimensional problems, such as the resonant radiation pressure on neutral particles in a waveguide~\cite{Gomez-Medina:2001}, strong optical interactions between particles~\cite{Gomez-Medina:2004}, resonance cooling of polarizable particles~\cite{Szirmai:2007}, and low-light-level optical interactions with atomic vapor in fiber~\cite{Ghosh:2006}.

J.-T. Shen acknowledges the informative discussions with J. Shin at Stanford and K. Srinivasan at NIST. S. Fan acknowledges financial support by the David and Lucile Packard Foundation.

\appendix
\section{Dissipations as Coupling to a Reservoir}\label{A:ReservoirCoupling}

In the effective Hamiltonian of Eq.~\eqref{E:Hamiltonian}, the dissipation rates of the cavity and the atom are characterized by $1/\tau_c$ and $1/\tau_a$, respectively. Here we show explicitly how these intrinsic dissipation rates emerge as the system $S$ couples to a reservoir $R$. To illustrate the physics, we will use as an example a system that consists of an atom in a waveguide, in which the atom also couples to a reservoir, to derive the effective Hamiltonian and to show that coupling to the reservoir yields a damped atom. The same procedures carry through for the cavity-atom case as well.

The Hamiltonian of the composite system $S\bigoplus R$ is $H\equiv H_S + H_R + H_{SR}$:
\begin{align}
H_{S}/\hbar & \equiv \int dx \,c_R^{\dagger}(x) \left(\omega_0-i v_g \frac{\partial}{\partial x}\right) c_R(x) +  \int dx \,c_L^{\dagger}(x) \left(\omega_0+i v_g \frac{\partial}{\partial x}\right) c_L(x)\notag\\
&+\int dx \,V \delta(x) \left[\sigma_{+}c_R(x) + c_R^{\dagger}(x) \sigma_{-} + \sigma_{+}c_L(x) + c_L^{\dagger}(x) \sigma_{-}\right]\notag\\
&+\Omega_e a_e^{\dagger}a_e +\Omega_g a_g^{\dagger}a_g,\\
H_R/\hbar &\equiv \sum_j \omega_j r_j^{\dagger}r_j,\\
H_{SR}/\hbar &\equiv \sum_j\left(\eta^{*}_j r_j^{\dagger}\sigma_{-} + \eta_j \sigma_+ r_j\right).
\end{align}

$H_S$ is the Hamiltonian of the system $S$ of coupled waveguide-atom. The reservoir $R$ is modeled as a collection of harmonic oscillators with frequencies $\omega_j$, and with the corresponding creation (annihilation) operators $r_{j}^{\dagger}$ ($r_{j}$).  $H_{SR}$ describes the interactions between the atom and the reservoir. 

An arbitrary time-independent one-excitation state in $S\bigoplus R$ is given by
\begin{align}\label{E:SRInteractingEigenstate}
|\epsilon^{+}\rangle_{S\bigoplus R} = & \int dx \left[\phi_R (x) c_R^{\dagger}(x) + \phi_L (x) c_L^{\dagger}(x)\right]|\emptyset\rangle + e_a \sigma_{+}|\emptyset\rangle + \sum_j e_j r^{\dagger}_j |\emptyset\rangle,
\end{align} which corresponds to the case that the atom was initially at the ground state and all oscillators were not excited when the incoming photon was at $-\infty$.
The Schr\"odinger equation
\begin{equation}\label{E:SRSchrodinger}
H|\epsilon^+\rangle_{S\bigoplus R} =\hbar \epsilon|\epsilon^+\rangle_{S\bigoplus R},
\end{equation}with $\epsilon = \omega +\Omega_g$, yields the equations of motion:
\begin{align}
\left(-i v_g \frac{\partial}{\partial x}\phi_R(x)\right) + eV\delta(x) &=\left(\epsilon-\omega_0 -\Omega_g\right) \phi_R(x),\label{E:SREofM1}\\
\left(+i v_g \frac{\partial}{\partial x}\phi_L(x)\right) + eV\delta(x) &=\left(\epsilon-\omega_0-\Omega_g\right) \phi_L(x),\label{E:SREofM2}\\
\Omega e_a + V \phi_R(0) + V\phi_L(0) +\sum_j \eta_j e_j &=\left(\epsilon-\Omega_g\right) e_a,\label{E:SREofM3}\\
\omega_j e_j + e_a \eta_j^* &=\left(\epsilon -\Omega_g\right) e_j.\label{E:SREofM4}
\end{align}
From Eq.~\eqref{E:SREofM4}:
\begin{equation}
e_j =\frac{\eta_j^*}{\omega-\omega_j} e_a,\label{E:SREofM4_2}
\end{equation}where $\omega = \epsilon -\Omega_g$ is the photon frequency. 

Plug Eq.~\eqref{E:SREofM4_2} into Eq.~\eqref{E:SREofM3}:
\begin{equation}\label{E:SREofM3_2}
\Omega e_a + V \phi_R(0) + V\phi_L(0) +\left(\sum_j \frac{{|\eta_j|}^2}{\omega -\omega_j}\right) e_a =\omega e_a.
\end{equation}Note that
\begin{align}\label{E:GammaDef}
\sum_j \frac{{|\eta_j|}^2}{\omega -\omega_j} &=\int d\omega'\, g(\omega')\frac{{|\kappa(\omega')|}^2}{\omega-\omega'+i\delta}\notag\\
&= P\int d\omega'\, g(\omega') \frac{{|\kappa(\omega')|}^2}{\omega-\omega'} - i \pi g(\omega){|\kappa(\omega)|}^2\notag\\
&\equiv \Delta - i {\gamma},
\end{align}where $g(\omega')$ is the density of states of the reservoir, and $P$ denotes the Cauchy principal value.
\begin{equation}
\Delta \equiv \int d\omega'\, g(\omega') P\frac{{|\eta(\omega')|}^2}{\omega-\omega'},
\end{equation}is the frequency  shift of the atom that is due to atom-reservoir coupling, and is analogous to the Lamb shift~\cite{Carmichael:2003};
\begin{equation}
\gamma \equiv \pi g(\omega){|\eta(\omega)|}^2
\end{equation}is the damping rate of the \emph{amplitude} due to atom-reservoir coupling. Note both $\Delta$ and $\gamma$  are functions of $\omega$.

Thus, Eq.~\eqref{E:SREofM3_2} becomes
\begin{equation}\label{E:SREofM3_3}
\left(\Omega+\Delta -i\gamma\right) e_a + V \phi_R(0) + V\phi_L(0) =\omega e_a,
\end{equation}and therefore
\begin{equation}
e_a =\frac{V \phi_R(0) + V\phi_L(0)}{\omega-\left(\Omega+\Delta\right)+i\gamma}.
\end{equation}When the coupling to reservoir is weak, one expects the real part of the pole $\omega$ of $e_a$ to be close to $\Omega$, and the corrections to be small such that one can evaluate $\Delta$ and $\gamma$ at $\omega=\Omega$:
\begin{align}
\int d\omega'\, g(\omega') P\frac{{|\eta(\omega')|}^2}{\omega-\omega'} \simeq \int d\omega'\, g(\omega') P\frac{{|\eta(\omega')|}^2}{\Omega-\omega'} &\equiv \Delta_a,\notag\\
\pi g(\omega){|\eta(\omega)|}^2\simeq \pi g(\Omega){|\eta(\Omega)|}^2 &\equiv \gamma_a,
\end{align}and finally
\begin{equation}\label{E:SREofM3_4}
\left(\Omega+\Delta_a -i\gamma_a\right) e_a + V \phi_R(0) + V\phi_L(0) =\omega e_a,
\end{equation}which is equivalent to the substitution $\Omega\rightarrow\Omega +\Delta_a -i\gamma_a$.

Using these expressions, one can show immediately that the three equations of motion that have only dynamic variables describing the system $S$, Eq.~\eqref{E:SREofM1}, \eqref{E:SREofM2}, and \eqref{E:SREofM3_4}, can be obtained from the Schr\"{o}dinger's equation
\begin{equation}
H_{\text{eff}} |\epsilon^{+}\rangle =\hbar \epsilon |\epsilon^{+}\rangle,
\end{equation}with the effective Hamiltonian
\begin{align}
H_{\text{eff}}/\hbar & \equiv \int dx \,c_R^{\dagger}(x) \left(\omega_0-i v_g \frac{\partial}{\partial x}\right) c_R(x) +  \int dx \,c_L^{\dagger}(x) \left(\omega_0+i v_g \frac{\partial}{\partial x}\right) c_L(x)\notag\\
&+\int dx \,V \delta(x) \left[\sigma_{+}c_R(x) + c_R^{\dagger}(x) \sigma_{-} + \sigma_{+}c_L(x) + c_L^{\dagger}(x) \sigma_{-}\right]\notag\\
&+\left(\Omega_e -i\frac{1}{\tau_a}\right) a_e^{\dagger}a_e +\Omega_g a_g^{\dagger}a_g,
\end{align}where $\gamma_a\equiv \frac{1}{\tau_a}$, and we have absorbed $\Delta_a$ into $\Omega_e$. The state is entirely in $S$ and is given by
\begin{equation}
|\epsilon^{+}\rangle =  \int dx \left[\phi_R (x) c_R^{\dagger}(x) + \phi_L (x) c_L^{\dagger}(x)\right]|\emptyset\rangle + e_a \sigma_{+}|\emptyset\rangle.
\end{equation}
This approach thus yields a damped atom.

\pagebreak
\newpage

\pagebreak
\newpage

\pagebreak
\newpage

\pagebreak
\newpage

\pagebreak
\newpage

\pagebreak
\newpage

\end{document}